\documentclass[12pt]{article}
\usepackage[T1]{fontenc}
\usepackage[latin9]{inputenc}
\usepackage[letterpaper]{geometry}
\usepackage{amstext}
\usepackage{graphicx}
\usepackage{setspace}
\usepackage{esint}
\makeatletter

\newcommand{\lyxaddress}[1]{
\par {\raggedright #1
\vspace{1.4em}
\noindent\par}
}

\makeatother

\begin{document}

\title{
\begin{flushright}
 \small MZ-TH/10-09
\end{flushright}
\vspace{2cm}
Polyakov Effective Action from Functional Renormalization Group Equation}

\author{Alessandro Codello%
\thanks{email: codello@thep.physik.uni-mainz.de%
}}

\maketitle

\lyxaddress{\begin{center}
Institut f\"ur Physik, Johannes Gutenberg-Universit\"at, Mainz\\
Staudingerweg 7, D-55099 Mainz, Germany
\par\end{center}}
\begin{abstract}
We discuss the Polyakov effective action for a minimally coupled scalar
field on a two dimensional curved space by considering a non-local
covariant truncation of the effective average action. We derive the
flow equation for the form factor in $\int\sqrt{g}R\, c_{k}(\Delta)R$,
and we show how the standard result is obtained when we integrate
the flow from the ultraviolet to the infrared.
\end{abstract}

\section{Introduction}

In recent years several groups have investigated the possibility that
a quantum theory of gravity can be constructed along the lines of
Asymptotic Safety \cite{Reuter_1998,Weinberg_1979,Codello_Percacci_Rahmede_2009,Litim_2004,Benedetti_Machado_Saueressig_2008}.
In this scenario metric degrees of freedom can still be used to construct
quantum gravity if a non trivial (other than Gaussian) fixed point
of the renormalization group (RG) flow with finite dimensional ultraviolet
(UV) critical surface can be found. Only after the introduction, in
this context, of an exact RG flow equation \cite{Reuter_1998} it
has become possible to properly investigate this issue. The evidence
for the existence of such a fixed point has gradually improved but
still much work has to be done to firmly establish this scenario and
to be able to make reliable predictions, especially at the IR scales.
Present day truncations have to be extended mainly in the direction
of including an infinite number of terms \cite{Codello_Percacci_Rahmede_2008,Machado_Saueressig_2008}
and to include non-local invariants. Actually the two issues are related
as to study non-local terms it is necessary to consider the RG flow
of functions of the covariant Laplacian. As a first step in this direction,
we will show how it is possible to write the flow equation for a,
possibly non-local, structure function in the simple case of the effective
average action for a minimally coupled scalar field on a two dimensional
manifold. This analysis is worth doing also because we already know
the answer: by integrating the conformal anomaly, Polyakov \cite{Polyakov_1981}
was able to calculate the effective action for this system. This is
one of the few cases in which the full non-local structure of the
effective action in known exactly. Reproducing Polyakov's result is
thus an important test for the exact RG flow equation approach, as
an example of a complete RG trajectory \cite{Manrique_Reuter_2009a},
and as an essential first step towards understanding more complicated
non-local invariants. 

The exact RG flow equation is a non-linear functional differential
equation that, in the simplest cases, involves a functional trace
over functions of the covariant Laplacian. The computation of these
traces is a technically rather difficult task, actually it is one
of the major obstacles in the development of the field. Heat kernel
techniques are intensively used. To be able to handle non-local terms
in truncations of the effective average action calls for the use of
the non-local heat kernel expansion developed by Barvinsky and Vilkovisky
\cite{Barvinsky_Vilkovinsky_1990}.

In the second section of the paper we briefly review the standard
derivation of the conformal anomaly and of the Polyakov effective
action. In the third we introduce the exact RG flow equation for a
minimally coupled scalar field on a $d$-dimensional manifold. In
the fourth section we show how to derive the flow equation for a non-local
truncation of the effective average action, and we use it to rederive
Polyakov's result.

\section{Conformal anomaly and Polyakov effective action}

We consider a minimally coupled scalar field $\phi$ on a two dimensional
background manifold $(\mathcal{M},g)$ where $g$ is a Riemannian
metric. We will work in Euclidean signature. The classical action
(CA) is\begin{equation}
S[\phi,g]=\frac{1}{2}\int d^{2}x\sqrt{g}g^{\mu\nu}\partial_{\mu}\phi\partial_{\nu}\phi=\frac{1}{2}\int d^{2}x\sqrt{g}\phi\Delta\phi\,,\label{1}\end{equation}
where we introduced the Laplacian operator $\Delta$ acting on scalar
fields:\begin{equation}
\Delta\phi=-\frac{1}{\sqrt{g}}\partial_{\mu}\left(\sqrt{g}g^{\mu\nu}\partial_{\nu}\phi\right)\,.\label{1.01}\end{equation}

As the action (\ref{1}) is invariant under diffeomorphisms, it follows
from Noether's theorem that the (classical) energy momentum tensor,
defined by\begin{equation}
T^{\mu\nu}=\frac{2}{\sqrt{g}}\frac{\delta S[\phi,g]}{\delta g_{\mu\nu}}\,,\label{1.1}\end{equation}
is conserved, $\nabla_{\mu}T^{\mu\nu}=0$. In the case of the minimally
coupled scalar we readily find\[
T^{\mu\nu}=\frac{1}{2}g^{\mu\nu}\left(\partial\phi\right)^{2}-\partial^{\mu}\phi\partial^{\nu}\phi\,.\]
It is easy to check that the energy momentum tensor is traceless if
$d=2$. This follows from the fact that the action (\ref{1}) is additionally
conformal invariant in the special case of a two dimensional manifold.
Under a Weyl rescaling $g_{\mu\nu}\rightarrow e^{\sigma}g_{\mu\nu}$
the covariant Laplacian, in $d$ dimensions, transforms as\[
\Delta\phi\rightarrow e^{-\sigma}\left[\Delta\phi-\left(\frac{d}{2}-1\right)g^{\mu\nu}\partial_{\mu}\sigma\partial_{\nu}\phi\right]\,,\]
and we see that only in $d=2$ the term $\sqrt{g}\Delta$ and the
action (\ref{1}) are invariant.

The generating functional of correlation functions for a scalar field
on the manifold $(\mathcal{M},g)$ is defined as usual by \begin{equation}
Z[J,g]=\int D_{g}\phi\,\exp\left(-S[\phi,g]+\int d^{d}x\sqrt{g}J\phi\right)\,,\label{2}\end{equation}
where $J$ is a scalar source field. Following \cite{Mottola_1995}
the functional integral measure can be defined by the requirements\begin{equation}
\int D_{g}\phi\, e^{-(\phi,\phi)_{g}}=1\qquad\qquad(\phi,\phi)_{g}=\frac{1}{2}\int d^{d}x\sqrt{g}\phi^{2}\label{2.1}\end{equation}
and is invariant under diffeomorphisms.

To find the expectation value of the energy momentum tensor we take
a functional derivative of (\ref{2}) with respect to the metric at
$J=0$:\begin{equation}
\frac{\delta Z[0,g]}{\delta g_{\mu\nu}}=-\int D_{g}\phi\,\frac{\delta S[\phi,g]}{\delta g_{\mu\nu}}\, e^{-S[\phi,g]}=-Z[0,g]\frac{\sqrt{g}}{2}\left\langle T^{\mu\nu}\right\rangle \,.\label{3}\end{equation}
Note that we did not differentiate the measure since we assumed it
is invariant under diffeomorphisms and has thus zero variation with
respect to the metric. We are assuming that there are no diffeomorphisms
anomalies. Recalling the definition of the generating functional of
connected correlation functions\[
W[J,g]=\log Z[J,g]\,,\]
we can write\begin{equation}
\left\langle T^{\mu\nu}\right\rangle =-\frac{2}{\sqrt{g}}\frac{\delta W[0,g]}{\delta g_{\mu\nu}}\,.\label{3.1}\end{equation}
This is the quantum energy momentum tensor. We can rewrite it in terms
of the effective action (EA) which is the Legendre transform of $W[J,g]$
with respect to the first argument, \begin{equation}
\Gamma[\varphi,g]=\int d^{2}x\sqrt{g}J(\varphi)\varphi-W[J(\varphi),g]\,,\label{3.2}\end{equation}
where $J=J(\varphi)$ is to be obtained by solving $\frac{\delta W}{\delta J}=\varphi$
for $J$. Using (\ref{3.1}) and (\ref{3.2}) at $J=0$ gives\begin{equation}
\left\langle T^{\mu\nu}\right\rangle =\frac{2}{\sqrt{g}}\frac{\delta\Gamma[\varphi,g]}{\delta g_{\mu\nu}}\,.\label{4}\end{equation}
The quantum energy momentum tensor (\ref{4}) is the functional derivative,
with respect to the metric, of the EA, as the classical energy momentum
tensor is the functional derivative, with respect to the metric, of
the CA.

Symmetries of the CA imply relations between the correlation functions
which are described by Ward-Takahashi (WT) identities. To derive the
WT identities related to diffeomorphism invariance, we consider an
infinitesimal diffeomorphism,\[
\delta_{\epsilon}g_{\mu\nu}=\nabla_{\mu}\epsilon_{\nu}+\nabla_{\nu}\epsilon_{\mu}\,.\]
The EA transforms as\begin{equation}
\delta_{\epsilon}\Gamma[0,g]=\int d^{2}x\,\delta_{\epsilon}g_{\mu\nu}\frac{\delta\Gamma[0,g]}{\delta g_{\mu\nu}}=-\int d^{2}x\,\sqrt{g}\,\epsilon_{\nu}\nabla_{\mu}\left\langle T^{\mu\nu}\right\rangle \,,\label{4.5}\end{equation}
where we used (\ref{4}) and integrated by parts. Being the EA invariant
under reparametrizations of the metric $\delta_{\epsilon}\Gamma[0,g]=0$,
equation (\ref{4.5}) implies $\nabla_{\mu}\left\langle T^{\mu\nu}\right\rangle =0$.
In the case of Weyl rescalings the EA transforms as\begin{equation}
\delta_{\sigma}\Gamma[0,g]=\int d^{2}x\,\delta_{\sigma}g_{\mu\nu}\,\frac{\delta\Gamma[0,g]}{\delta g_{\mu\nu}}=\frac{1}{2}\int d^{2}x\,\sqrt{g}\,\delta\sigma\,\left\langle T_{\;\mu}^{\mu}\right\rangle \,.\label{5}\end{equation}
In this case we expect $\left\langle T_{\;\mu}^{\mu}\right\rangle \neq0$
because the functional measure we chose (\ref{2.1}) is not Weyl invariant.
To see this consider an infinitesimal Weyl rescaling $\delta_{\sigma}g_{\mu\nu}=g_{\mu\nu}\delta\sigma$.
The functional measure transforms as:\[
\delta_{\sigma}(\phi,\phi)_{g}=\frac{1}{2}\int d^{2}x\,\delta_{\sigma}\sqrt{g}\phi^{2}=\frac{1}{2}\int d^{2}x\,\sqrt{g}\delta\sigma\phi^{2}\neq0\,.\]
We thus expect an anomaly in the WT identity related to Weyl rescalings
(\ref{5}).

The functional integral (\ref{2}) with the CA (\ref{1}) is Gaussian.
Hence the exact EA is given by the one-loop formula\begin{eqnarray}
\Gamma[0,g] & = & S[0,g]+\frac{1}{2}\textrm{Tr}\log\, S^{(2)}[0,g]\nonumber \\
 & = & -\frac{1}{2}\int_{1/\Lambda^{2}}^{\infty}\frac{dt}{t}\textrm{Tr}\, e^{-t\Delta}\,.\label{6}\end{eqnarray}
 We have regularized the UV divergences of the parameter integral
introducing an UV cutoff $\Lambda$ and we used the simple relation
$S^{(2)}[0,g]=\Delta$. The EA is independent of the mean value of
the scalar field $\varphi$ and is a purely geometrical functional
constructed with invariants of the metric tensor $g$.

We will look for a possible Weyl anomaly by inserting equation (\ref{6})
into the WT relation (\ref{5}). Under a Weyl rescaling the EA changes
as\[
\delta_{\sigma}\Gamma[0,g]=-\frac{1}{2}\int_{1/\Lambda^{2}}^{\infty}\frac{dt}{t}\,\textrm{Tr}\,\delta_{\sigma}e^{-t\Delta}\,.\]
Some simple manipulations\[
\frac{1}{t}\textrm{Tr}\left[\delta_{\sigma}e^{-t\Delta}\right]=-\textrm{Tr}\left[\delta_{\sigma}\Delta\, e^{-t\Delta}\right]=\textrm{Tr}\left[\delta\sigma\,\Delta\, e^{-t\Delta}\right]=-\frac{d}{dt}\textrm{Tr}\left[\delta\sigma\, e^{-t\Delta}\right]\]
give\[
\delta_{\sigma}\Gamma[0,g]=\frac{1}{2}\textrm{Tr}\left[\delta\sigma\, e^{-\Delta/\Lambda^{2}}\right]\,.\]
The trace in the last expression can be calculated by the heat kernel
using the local expansion (\ref{HK_2}) derived in the appendix. We
find\begin{equation}
\textrm{Tr}\left[\delta\sigma\, e^{-\Delta/\Lambda^{2}}\right]=\frac{\Lambda^{2}}{4\pi}\int d^{2}x\sqrt{g}\,\delta\sigma+\frac{1}{24\pi}\int d^{2}x\sqrt{g}\, R\,\delta\sigma+O\left(\frac{1}{\Lambda^{2}}\right)\,.\label{6.1}\end{equation}
The first, divergent term is the Weyl variation of $\int\sqrt{g}$
(in $d=2$) and can be renormalized by introducing in $S[\phi,g]$
a conformal symmetry breaking {}``cosmological term'' $a_{\Lambda}\int\sqrt{g}$
and choosing $a_{\Lambda}=-\frac{\Lambda^{2}}{8\pi}$. The renormalized
cosmological constant is zero then. We cannot renormalize the second
term in equation (\ref{6.1}) because it is topological and thus its
Weyl variation is a total derivative:\[
\delta_{\sigma}\int d^{2}x(\sqrt{g}R)=\int d^{2}x\sqrt{g}\left[\delta\sigma R+\sqrt{g}(-R\delta\sigma+\Delta\delta\sigma)\right]=\int d^{2}x\sqrt{g}\Delta\delta\sigma\,.\]
Comparing with (\ref{5}) we find the two dimensional \emph{conformal
anomaly} induced by a minimally coupled scalar field:\begin{equation}
\left\langle T_{\;\mu}^{\mu}\right\rangle =-\frac{R}{24\pi}\,.\label{7}\end{equation}
More generally, if we consider $N_{S}$ scalar fields and $N_{F}$
Dirac fermions we find\[
\left\langle T_{\;\mu}^{\mu}\right\rangle =-\frac{c}{24\pi}R\qquad\qquad c=N_{S}+N_{F}\,,\]
where $c$ is the conformal anomaly coefficient.

Using the fact that all metrics on a two dimensional manifold are
locally conformally flat, Polyakov \cite{Polyakov_1981} showed that
it is possible to calculate the full EA by functionally integrating
the conformal anomaly. To do this we write the metric as conformally
equivalent to the flat metric: $g_{\mu\nu}=e^{\sigma}\delta_{\mu\nu}$.
The Ricci scalar becomes\begin{equation}
R=-e^{-\sigma}\partial^{2}\sigma\label{7.1}\end{equation}
 and using the anomaly equation (\ref{7}) together with equation
(\ref{4}) we find \[
\frac{\delta\Gamma[0,e^{\sigma}\delta]}{\delta\sigma}=-\frac{1}{48\pi}(-\partial^{2})\sigma\,.\]
This relation is trivially integrated to give\[
\Gamma[0,e^{\sigma}\delta]=-\frac{1}{96\pi}\int d^{2}x\,\sigma(-\partial^{2})\sigma+\textrm{const}\,,\]
where the constant can be set to be zero. Using $\Delta=-e^{-\sigma}\partial^{2}$,
and equation (\ref{7.1}) in the reverse way as before, we finally
find\begin{equation}
\Gamma[0,g]=-\frac{1}{96\pi}\int d^{2}x\sqrt{g}\, R\frac{1}{\Delta}R\,.\label{8}\end{equation}
This is the Polyakov EA for a minimally coupled scalar field on the
manifold \emph{$(\mathcal{M},g)$ }\cite{Polyakov_1981}\emph{.} Considering
that $\sqrt{g_{x}}\Delta_{x}G_{xy}=\delta_{xy}$%
\footnote{Here we use the compact notation $\delta_{xy}\equiv\delta^{(2)}(x-y)$,
$G_{xy}\equiv G(x-y)$ and so on.%
} we actually have to write\begin{equation}
\Gamma[0,g]=-\frac{1}{96\pi}\int d^{2}xd^{2}y\sqrt{g_{x}}\sqrt{g_{y}}\, R_{x}G_{xy}R_{y}\,.\label{9}\end{equation}
Note that EA (\ref{8}) is non-local since it involves the two point
correlation function $G_{xy}=(\Delta^{-1})_{xy}$ evaluated at different
points of $\mathcal{M}$, but it is analytical in the curvatures.\\
Having at our disposal the complete quantum effective action (\ref{8})
we can, for example, calculate the quantum energy momentum tensor
using equation (\ref{4}). After a few manipulations we find \cite{Mukhanov_Winitzki}
\begin{eqnarray}
\left\langle T^{\mu\nu}\right\rangle  & = & \frac{1}{48\pi}\left[-2\nabla^{\mu}\nabla^{\nu}\frac{1}{\Delta}R-\left(\nabla^{\mu}\frac{1}{\Delta}R\right)\left(\nabla^{\nu}\frac{1}{\Delta}R\right)+\right.\nonumber \\
 &  & \left.-2g^{\mu\nu}R+\frac{1}{2}g^{\mu\nu}\left(\nabla^{\alpha}\frac{1}{\Delta}R\right)\left(\nabla_{\alpha}\frac{1}{\Delta}R\right)\right]\,.\label{10}\end{eqnarray}
If we take the trace of equation (\ref{10}) we can check that the
conformal anomaly (\ref{7}) is correctly reproduced. Expression (\ref{10})
is exact, non-local and contains polarization as well as particle
production effects; the Hawking radiation in $d=2$ can be derived
starting from this expression for the energy momentum tensor \cite{Mukhanov_Winitzki}.
Note that in the limit of flat space we have $\left\langle T^{\mu\nu}\right\rangle =0$
since the scalar field becomes free.

\section{Effective average action}

The effective average action (EAA) $\Gamma_{k}[\varphi,g]$ is a functional
that interpolates smoothly between the CA and the EA. We will construct
the EAA for a scalar on the $d$-dimensional manifold $(g,\mathcal{M})$.
More details on the EAA on curved backgrounds and in presence of quantized
gravity can be found in \cite{Codello_Percacci_Rahmede_2009}.

Starting from the functional integral (\ref{2}) we add to the CA
an infrared (IR) cutoff or {}``regulator'' term $\Delta S_{k}[\phi,g]$
of the form\[
\Delta S_{k}[\phi,g]=\frac{1}{2}\int d^{d}x\sqrt{g}\phi\, R_{k}(\Delta)\,\phi\,,\]
where the kernel $R_{k}(\Delta)$ is chosen so to suppress the field
modes $\phi_{n}$, eigenfunctions of the Laplacian $\Delta\phi_{n}=\lambda_{n}\phi_{n}$,
with eigenvalues smaller than the cutoff scale $\lambda_{n}<k^{2}$.
The functional form of $R_{k}(z)$ is arbitrary except for the requirements
that it should be a monotonically decreasing function in both $z$
and $k$, that $R_{k}(z)\rightarrow0$ for $z\gg k^{2}$ and that
$R_{k}(z)\rightarrow k^{2}$ for $z\ll k^{2}$.

The scale dependent generating functional of correlation functions
that generalizes equation (\ref{2}) is \begin{equation}
Z_{k}[J,g]=\int D_{g}\phi\,\exp\left(-S[\phi,g]-\Delta S_{k}[\phi,g]+\int\sqrt{g}J\phi\right)\,.\label{10.1}\end{equation}
The scale dependent EA, referred to as the effective average action
(EAA) is defined by the relation\begin{equation}
\Gamma_{k}[\varphi,g]+\Delta S_{k}[\varphi,g]=\int\sqrt{g}J(\varphi)\varphi-W_{k}[J(\varphi),g]\,.\label{10.2}\end{equation}
Note that the Legendre transform of the generating functional of the
connected correlation functions is $\Gamma_{k}[\varphi,g]+\Delta S_{k}[\varphi,g]$
and not the EAA; thus $\Gamma_{k}[\varphi,g]$ need not to be a convex
functional for non zero $k$.

In the limits $k\rightarrow0$, and $k\rightarrow\Lambda$ the EAA
approaches, respectively, the EA and the bare CA \cite{Manrique_Reuter_2009a}:\[
\lim_{k\rightarrow0}\Gamma_{k}[\varphi,g]=\Gamma[\varphi,g]\qquad\qquad\lim_{k\rightarrow\Lambda}\Gamma_{k}[\varphi,g]=S[\varphi,g]\,.\]
The EAA and its RG flow offer a different approach to quantization.
In theory space, the space of {}``all'' action functionals, the
bare action represents the UV starting point of a RG trajectory which
reaches the EA in the IR. The integration of successive modes is done
step by step when lowering the cutoff scale $k$.

There is a considerable freedom in the choice of the functional form
of the regulator kernel. We will use the following three examples
here, the {}``optimized'' \cite{Litim_2001}, the exponential, and
the mass-type cutoff, respectively:\begin{eqnarray}
R_{k}^{opt}(z) & = & (k^{2}-z)\theta(k^{2}-z)\nonumber \\
R_{k}^{exp}(z) & = & \frac{z}{e^{z/k^{2}}-1}\nonumber \\
R_{k}^{mass}(z) & = & k^{2}\,.\label{11}\end{eqnarray}
The mass cutoff is not properly a cutoff because for $z\gg k^{2}$
it does not go to zero. Hence it does not guarantee the UV finiteness
of the flow. Still it is useful since it often allows for analytical
computations of UV finite quantities. 

In the case we will study in this paper, where the bare propagator
goes like $\frac{1}{p^{2}}\equiv\frac{1}{z}$, it is easy to see how
the cutoff acts as an IR regulator. The regularized propagator $\frac{1}{z+R_{k}(z)}$
is shown in Figure 1 for the three different cutoff shapes in (\ref{11})
plotted together with the bare one. For modes of momentum eigenvalues
greater then the RG scale $z\gg k^{2}$ the propagation is unaffected
while starting at the cutoff scale their propagation is successively
suppressed as if they were massive particles of mass $k$.%
\begin{figure}
\centering{}\includegraphics{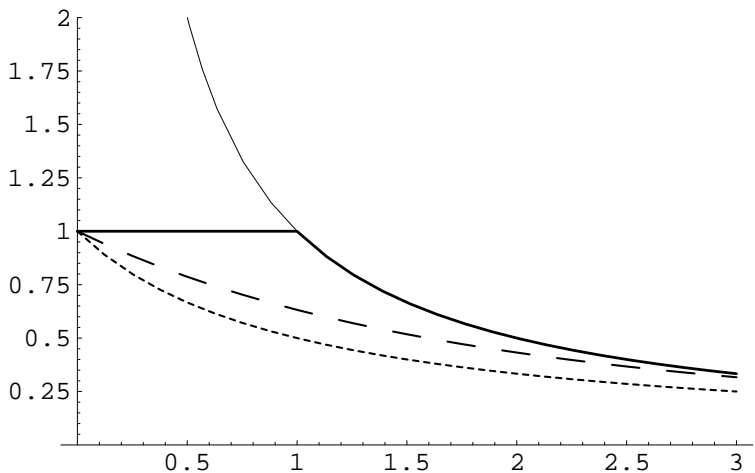}\caption{Dimensionless bare propagator (continuous) together with regularized
propagators using optimized cutoff (thick), exponential cutoff (long
dashed) and mass-type cutoff (short dashed) plotted as functions of
$z/k^{2}$.}

\end{figure}

It is possible to derive an exact RG flow equation for the EAA by
differentiating equation (\ref{10.2}) with respect to the {}``RG
time'' $t=\log k/k_{0}$ ($k_{0}$ is a reference scale) as first
done in \cite{Wetterich_1993} for the case of a scalar field on a
$d$ dimensional flat background. The flow equation for the situation
we are considering of a $d$ dimensional scalar field on a curved
background \cite{Codello_Percacci_Rahmede_2009} reads\begin{equation}
\partial_{t}\Gamma_{k}[\varphi,g]=\frac{1}{2}\mathrm{Tr}\frac{\partial_{t}R_{k}[g]}{\Gamma_{k}^{(2,0)}[\varphi,g]+R_{k}[g]}\,.\label{erge}\end{equation}
Here $\Gamma_{k}^{(2,0)}[\varphi,g]$ is the second functional derivative
of the EAA with respect to $\varphi$. The trace in (\ref{erge})
is a sum over the eigenvalues of the covariant Laplacian and is usually
evaluated using heat kernel techniques as explained in the Appendix.
Equation (\ref{erge}) is exact and is both UV and IR finite.

The flow equation (\ref{erge}) can be seen as a RG improvement of
the one-loop effective action derived from the bare action $S[\varphi,g]+\Delta S_{k}[\varphi,g]$.
Using the definition of the EAA (\ref{10.2}) and the standard one-loop
relation for the EA we find\begin{equation}
\Gamma_{k}[\varphi,g]=S[\varphi,g]+\frac{1}{2}\textrm{Tr}\log\left(S^{(2,0)}[\varphi,g]+R_{k}[g]\right)\,,\label{one_loop}\end{equation}
Differentiating with respect to the RG time gives\begin{equation}
\partial_{t}\Gamma_{k}[\varphi,g]=\frac{1}{2}\mathrm{Tr}\frac{\partial_{t}R_{k}[g]}{S^{(2,0)}[\varphi,g]+R_{k}[g]}\,.\label{erge_one_loop}\end{equation}
If now we {}``RG improve'' equation (\ref{erge_one_loop}) in the
sense of replacing in the right side the Hessian of the bare action
with the Hessian of the EAA we recover the exact flow equation (\ref{erge}).

\section{Flow equations}

For a minimally coupled scalar field the one-loop EAA (\ref{one_loop})
is already exact and it is thus enough to consider the flow equation
(\ref{erge_one_loop}). The right hand side of (\ref{erge_one_loop})
is independent of $\varphi$ so the flow will not generate any non
trivial dependence of the EAA on $\varphi$. We will thus concentrate
on the flow of $\Gamma_{k}[0,g]$ which is the non trivial part of
the EAA. 

The Hessian of the bare action (\ref{1}) is just the covariant Laplacian
$S^{(2,0)}[0,g]=\Delta$ and equation (\ref{erge_one_loop}) reduces
to \begin{equation}
\partial_{t}\Gamma_{k}[0,g]=\frac{1}{2}\textrm{Tr}\frac{\partial_{t}R_{k}(\Delta)}{\Delta+R_{k}(\Delta)}\,,\label{12}\end{equation}
We can evaluate the trace in equation (\ref{12}) using the technology
developed in the Appendix to which the reader could turn at this point.
Defining the function $h(z)=\frac{\partial_{t}R_{k}(z)}{z+R_{k}(z)}$
and using the heat kernel non-local expansion (\ref{HK_6}) in equation
(\ref{HK_1.1}) from the Appendix we find\begin{eqnarray}
\textrm{Tr}\, h(\Delta) & = & \frac{1}{4\pi}Q_{1}[h]\int d^{2}x\sqrt{g}+\frac{1}{24\pi}Q_{0}[h]\int d^{2}x\sqrt{g}R+\nonumber \\
 &  & +\frac{1}{4\pi}\int d^{2}x\sqrt{g}R\left[\int_{0}^{\infty}ds\,\tilde{h}(s)\, s\, f_{R}(s\Delta)\right]R+O(R^{3})\,,\label{13}\end{eqnarray}
Here $\tilde{h}(s)$ is the Laplace transform of the function $h(x)$
and the $Q$-functional are defined in (\ref{HK_8}). To make progress
we need to devise a truncation of the EAA to insert into the left
hand side of (\ref{12}). We will consider an ansatz where the EAA
is local in the curvature (analytical in the metric) but non-local
in the covariant momentum squared, i.e. in $\Delta$. We are lead
to the following truncation ansatz which comprises the first terms
of the curvature expansion:\begin{equation}
\Gamma_{k}[0,g]=\int d^{2}x\sqrt{g}\left(a_{k}+b_{k}R+R\, c_{k}(\Delta)R\right)+O(R^{3})\,.\label{14}\end{equation}
Here $c_{k}(x)$ is any function of the covariant Laplacian. We are
working in two dimensions so this is the only structure function at
second order in the curvature to be considered.

By comparing equation (\ref{13}) to (\ref{14}), the beta functions
for the first two couplings in (\ref{14}) are immediately found:\begin{eqnarray}
\partial_{t}a_{k} & = & \frac{1}{8\pi}Q_{1}[h]\nonumber \\
\partial_{t}b_{k} & = & \frac{1}{48\pi}Q_{0}[h]\,.\label{flow_0}\end{eqnarray}
From the curvature squared term we find the flow of the non-local
structure function\begin{equation}
\partial_{t}c_{k}(x)=\frac{1}{8\pi}\int_{0}^{\infty}ds\,\tilde{h}_{k}(s)\, s\, f_{R}(sx)\,.\label{flow_1}\end{equation}
If we now insert the explicit form of the heat kernel structure function
$f_{R}(x)$ from equation (\ref{HK_7}) of the Appendix, written in
terms of the basic parameter integral (\ref{HK_5}), and use relation
(\ref{HK_8}), we find:\begin{eqnarray}
8\pi\,\partial_{t}c_{k}(x) & = & \frac{1}{32}\int_{0}^{1}d\xi\, Q_{-1}\left[h(z+x\xi(1-\xi))\right]+\nonumber \\
 &  & +\frac{1}{8x}\int_{0}^{1}d\xi\, Q_{0}\left[h(z+x\xi(1-\xi))\right]-\frac{1}{16x}Q_{0}[h]+\nonumber \\
 &  & +\frac{3}{8x^{2}}\int_{0}^{1}d\xi\, Q_{1}\left[h(z+x\xi(1-\xi))\right]-\frac{3}{8x^{2}}Q_{1}[h]\,.\label{flow_2}\end{eqnarray}
In the last equation the dummy index $z$ is shown to indicate that
the $Q$-functionals are to be evaluated at the shifted point $z+x\xi(1-\xi)$.
The next step is to use (\ref{HK_9}) to calculate the $Q$-functionals:\begin{eqnarray}
8\pi\,\partial_{t}c_{k}(x) & = & -\frac{1}{32}\int_{0}^{1}d\xi\, h'\left(x\xi(1-\xi)\right)+\frac{1}{8x}\int_{0}^{1}d\xi\, h\left(x\xi(1-\xi)\right)+\nonumber \\
 &  & -\frac{1}{16x}h(0)-\frac{3}{8x^{2}}\int_{0}^{1}d\xi\,\int_{0}^{x\xi(1-\xi)}dz\, h\left(z\right)\,.\label{flow_exp}\end{eqnarray}
Note that we combined the last two terms of equation (\ref{flow_2})
into a single $z$ integral.

Equation (\ref{flow_exp}) is the explicit flow equation for the structure
function $c_{k}(x)$. It should be possible to integrate equation
(\ref{flow_exp}) from the UV to the IR scale to recover the Polyakov
EA (\ref{8}) without specifying the cutoff shape function $R_{k}(z)$.
Here we will show how this can be done by explicitly using the cutoff
shapes in (\ref{11}).

First we use the {}``optimized'' cutoff to evaluate the beta functions
(\ref{flow_0}):\begin{eqnarray}
\partial_{t}a_{k} & = & \frac{k^{2}}{4\pi}\nonumber \\
\partial_{t}b_{k} & = & \frac{1}{24\pi}\,.\label{flow_3}\end{eqnarray}
After collecting the overall power $k^{-2}$ and writing the parameter
integrals in terms of the dimensionless variable $x/k^{2}$, the flow
equation (\ref{flow_exp}) can be rewritten in the form \begin{equation}
\partial_{t}c_{k}(x)=\frac{1}{8\pi\, k^{2}}f\left(\frac{x}{k^{2}}\right)\,.\label{flow_4}\end{equation}
The function $f(u)$ depends explicitly on the cutoff shape function
used. In the case of the optimized and mass cutoffs we find after
some elementary integrations, respectively,\begin{eqnarray}
f_{opt}(u) & = & \frac{1}{8u}\left[\sqrt{\frac{u}{u-4}}-\frac{u+4}{u}\sqrt{\frac{u-4}{u}}\right]\theta(u-4)\nonumber \\
f_{mass}(u) & = & \frac{\sqrt{u(u+4)}(u+6)+8(u+3)\,\textrm{artanh}\sqrt{\frac{u}{u+4}}}{(u+4)^{3/2}u^{5/2}}\,.\label{flow_f}\end{eqnarray}
The parameter integrals in equation (\ref{flow_exp}) cannot be evaluated
analytically for the exponential cutoff, but this can still be done
numerically. The functions $f(u)$ evaluated for the three different
cutoffs are plotted in Figure 2%
\begin{figure}
\begin{centering}
\includegraphics{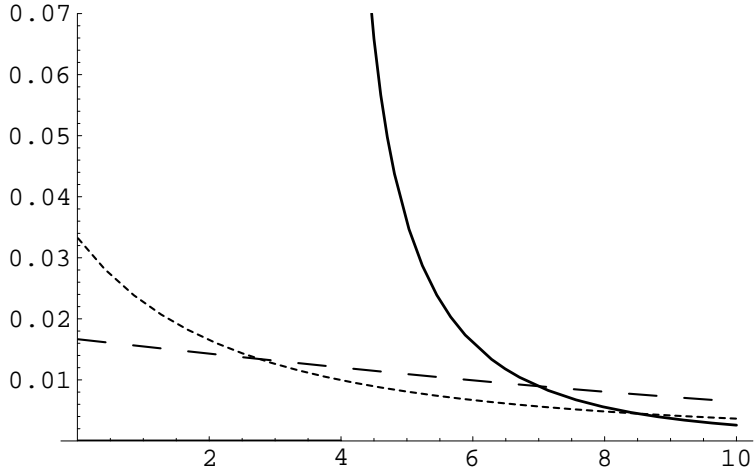}
\par\end{centering}

\caption{The function $f(u)$ evaluated using the exponential cutoff (long
dashed), the mass cutoff (short dashed) and the optimized cutoff (thick).
Note that all three functions are analytic around the origin and that
$f_{opt}(u)$ develops a pole at $u=4$.}

\end{figure}
. Note that they are all analytic in a neighborhood of the origin,
$f_{opt}(u)$ is even zero in the entire interval $[0,4)$.

If we were to interpret $f(u)$ as a power series in $u$ about $u=0$,
it follows that we have a non zero running of local terms of the form
$c_{k}^{(n)}\int\sqrt{g}R\Delta^{n}R$ only for the exponential and
the mass cutoff. For example, we can expand for small $u$\[
f_{mass}(u)=\frac{1}{30}-\frac{u}{70}+\frac{u^{2}}{210}+O(u^{3})\,,\]
and read off the resulting beta functions for the couplings $c_{k}^{(n)}$
in the mass cutoff case. For the optimized cutoff none of the couplings
$c_{k}^{(n)}$ has a non-zero beta function. Any finite truncation
of the EAA containing some of the couplings $c_{k}^{(n)}$ will never
reproduce the correct IR behavior and will lead to IR divergences.
More importantly, the running of the couplings $c_{k}^{(-n)}$, $n>0$,
which multiply non-local terms involving inverse powers of $\Delta$
is zero for all three cutoff choices. In particular, the beta function
of the coupling $c_{k}^{(-1)}$ pertaining to the operator $\int\sqrt{g}R\frac{1}{\Delta}R$
is zero, even if this is the form the EAA is expected to reach at
$k=0$! We conclude that, at least for the cutoff shapes we considered,
to capture the non-local features of the EAA we need to consider the
running of the whole structure function.

We now integrate the flow equations from the UV scale $\Lambda$,
where we impose the initial conditions $\Gamma_{\Lambda}[\varphi,g]=S_{\Lambda}[\varphi,g]$,
to the IR scale $k$. We will see that imposing the initial conditions
not only selects which theory we are quantizing, but also implements
the renormalization conditions. In the limit $k\rightarrow0$ we will
find the full EA\@.

We start solving the differential equations (\ref{flow_3}). Integrating
from $k$ to $\Lambda$ gives\begin{eqnarray}
a_{k} & = & a_{\Lambda}-\frac{1}{4\pi}(\Lambda^{2}-k^{2})\nonumber \\
b_{k} & = & b_{\Lambda}-\frac{1}{24\pi}\log\frac{\Lambda}{k}\,.\label{int_flow_0}\end{eqnarray}
The coupling $a_{k}$ has to be renormalized, this can be done by
setting $a_{\Lambda}=\frac{\Lambda^{2}}{4\pi}$ so that the renormalized
coupling vanishes in the IR, $a_{0}=0$. This preserves conformal
invariance of the EA.

Integrating the RG equation (\ref{flow_4}) of the structure function
gives\[
c_{k}(x)=c_{\Lambda}(x)-\frac{1}{8\pi}\int_{k}^{\Lambda}\frac{dk'}{k'^{3}}\, f\left(\frac{x}{k'^{2}}\right)\,.\]
If we use the variable $y=x/k^{2}$ we have $dk/k^{3}=-dy/2x$ and
we come to\begin{equation}
c_{k}(x)=c_{\Lambda}(x)-\frac{1}{16\pi x}\int_{x/\Lambda^{2}}^{x/k^{2}}dy\, f\left(y\right)\,.\label{int_flow_1}\end{equation}
If the integral in (\ref{int_flow_1}) is convergent at both the lower
and upper limit, it becomes a pure number in the limit $\Lambda\rightarrow\infty$
and $k\rightarrow0$. The functional form of $c_{0}(x)$ will be in
agreement with the Polyakov effective action (\ref{8}) if \[
\int_{0}^{\infty}dy\, f\left(y\right)=\frac{1}{6}\,.\]
This condition should be met for any cutoff choice. A simple integration
of (\ref{flow_f}) shows that this is so for the optimized and mass
cutoffs. A numerical evaluation shows that also the exponential cutoff
gives the desired result.

Imposing the boundary condition $c_{\infty}(x)=0$, we thus recover
Polyakov's non-local EA, equation (\ref{8}), as the result of the
integration of the flow:\[
c_{0}(x)=-\frac{1}{96\pi x}\,.\]

We are now in a position to write down the EAA, within truncation
(\ref{14}) and using the optimized cutoff (\ref{11}), at any given
scale $k$ we have:\begin{eqnarray*}
\Gamma_{k}[0,g] & = & \frac{k^{2}}{4\pi}\int d^{2}x\sqrt{g}-\frac{1}{24\pi}\log\frac{\Lambda}{k}\int d^{2}x\sqrt{g}R\;+\qquad\qquad\qquad\qquad\qquad\qquad\end{eqnarray*}
\begin{equation}
\qquad\qquad-\frac{1}{96\pi}\int d^{2}x\sqrt{g}R\left[\frac{\sqrt{\Delta/k^{2}-4}(\Delta/k^{2}+2)}{\Delta\left(\Delta/k^{2}\right)^{3/2}}\theta(\Delta/k^{2}-4)\right]R+O(R^{3})\,.\label{flow_final}\end{equation}
This relation is our main result. It shows how the EAA interpolates
smoothly between the classical action at the scale $k=\Lambda$ and
the EA at the scale $k=0$.

Note that in principle we still have to show that all higher terms
that would extend the truncation (\ref{14}) to higher curvature terms,
and are in principle present in the EAA, vanish at $k=0$. Only then
we would have completely recovered Polyakov's result. We shall not
embark on such a proof since this issue is special to two dimensions
and, contrary to the above discussion, does not generalize to higher
dimensions \cite{Satz_Mazzitelli_Codello_2010}.

\section{Conclusion}

In this paper we explained how the Polyakov effective action for a
minimally coupled scalar field on a curved two dimensional manifold
emerges within the functional RG approach. To do this we calculated
the RG flow of the structure function $c_{k}(\Delta)$ using the non-local
heat kernel expansion. We learned that in order to be able to recover,
at the IR scale, special non-local terms in the EAA, $\int\sqrt{g}R\,\frac{1}{\Delta}R$
in our example, it is necessary to include the running of the complete
structure function which allows for an arbitrary dependence on $\Delta$.
We also saw that, quite remarkably, individual non-local terms in
a Laurent series expansion, $\int\sqrt{g}R\Delta^{-n}R$, $n>0$,
have no RG running, even though the $k\rightarrow0$ limit of the
EAA is precisely of this type. This is an important observation in
view of future physical applications when we move on to consider quantized
gravity, especially in four dimensions. In that case the flow equations
for the corresponding structure functions will be a non-linear integro-differential
equation since the flow will no longer be one-loop like as in the
simple case treated in this paper. This strategy seems promising for
finding the IR completion of the RG flow of Quantum Einstein Gravity
which was computed within local truncations and found to terminate
at unphysical singularities in the IR \cite{Reuter_Saueressig_2002a}.

\section*{Acknowledgments}

I would like to thank M. Reuter, R. Percacci, F. Saueressig, E. Manrique,
D. Becker for careful reading the manuscript and A. Satz for useful
and stimulating discussion.

\appendix

\section{Non local Heat Kernel expansion}

In this appendix we review the basic relations related to the heat
kernel expansion. We will consider both the local and the non-local
expansion of the heat kernel for the covariant Laplacian (\ref{1.01})
on a general $d$-dimensional manifold.

The heat kernel $K_{xy}(s)$ satisfies the following differential
equation with boundary condition:\begin{equation}
\left(\partial_{s}+\Delta_{x}\right)K_{xy}(s)=0\qquad\qquad K_{xy}(0)=\delta_{xy}\,.\label{HK_1}\end{equation}
This equation describes the diffusion of a test particle on the manifold.
Here $s$ is the heat kernel {}``propertime'' parameter which is
related to the diffusion constant $D$ and to time $t$ as $s=Dt$.
We use the compact notation $K_{xy}(s)\equiv K(s;x,y)$ and $\delta_{xy}\equiv\delta^{(d)}(x-y)$.
Equation (\ref{HK_1}) is solved by $K_{xy}^{s}=e^{-s\Delta_{x}}\delta_{xy}$
and the trace of the heat kernel is thus equal to $\textrm{Tr}\, K(s)=\textrm{Tr}\, e^{-s\Delta}$.

The usefulness of the heat kernel expansion stems from the fact that
every functional trace of a function $h(\Delta)$ of the covariant
Laplacian can be related to it by a Laplace transform:\begin{equation}
\textrm{Tr}\, h(\Delta)=\int_{0}^{\infty}ds\,\tilde{h}(s)\,\textrm{Tr}\, e^{-s\Delta}\,.\label{HK_1.1}\end{equation}
$\tilde{h}(s)$ is the Laplace transform of $h(x)$. To compute such
a functional trace one just need to know the expansion of the heat
kernel trace to the desired accuracy.

For the trace there exists a standard asymptotic series in local curvature
polynomials,\begin{equation}
\textrm{Tr}\, K(s)=s^{-d/2}\left(B_{0}[\Delta]+sB_{2}[\Delta]+s^{2}B_{4}[\Delta]+...\right)\,,\label{HK_2}\end{equation}
where the $B_{2n}$ are the integrated heat kernel coefficients. They
are related to the un-integrated coefficients $b_{2n}$ by the relation
\[
B_{2n}[\Delta]=\frac{1}{(4\pi)^{d/2}}\int d^{d}x\sqrt{g}\,\textrm{tr}\, b_{2n}[\Delta](x)\,.\]
For the Laplacian (\ref{1.01}) operator $\Delta$ acting on scalars
the first few are\[
b_{0}[\Delta]=1\qquad\qquad b_{2}[\Delta]=\frac{R}{6}\]
\begin{equation}
b_{4}[\Delta]=\frac{1}{180}\left(R_{\mu\nu\alpha\beta}^{2}-R_{\mu\nu}^{2}+\frac{5}{2}R^{2}-6\nabla^{2}R\right)\,.\label{HK_3}\end{equation}
For the purposes of this paper we need a more sophisticated version
of the heat kernel expansion developed by Barvinsy and Vilkovinsky
\cite{Barvinsky_Vilkovinsky_1990} which includes an infinite number
of terms in the form of non-local structure factors. The generalized
non-local expansion that replaces (\ref{HK_2}) reads\begin{eqnarray}
\textrm{Tr}\, K(s) & = & \frac{1}{(4\pi s)^{d/2}}\int d^{d}x\sqrt{g}\left[1+s\frac{R}{6}+s^{2}R_{\mu\nu}f_{a}(s\Delta)R^{\mu\nu}+\right.\nonumber \\
 &  & \left.+s^{2}R\, f_{b}(s\Delta)R+O(R^{3})\right]\,,\label{HK_4}\end{eqnarray}
where $f_{a}(x)$ and $f_{b}(x)$ are two particular linear combinations
\cite{Barvinsky_Vilkovinsky_1990} of the basic structure factor \begin{equation}
f(x)=\int_{0}^{1}d\xi\, e^{-x\xi(1-\xi)}\,.\label{HK_5}\end{equation}
In equation (\ref{HK_4}) only two of the three possible curvature
square terms appear, the third one has been eliminated by using Bianchi's
identities and discarding boundary terms. For this reason the total
derivative term in the coefficient \textbf{$B_{4}$} is not present
and a straightforward series expansion of the structure functions
$f_{a}(x)$ and $f_{b}(x)$ will not give the same numerical coefficients
as in (\ref{HK_3}). This is because the form factor expansion (\ref{HK_4})
gives directly the integrated coefficients $B_{2n}$ while the standard
expansion gives the un-integrated coefficients $b_{2n}$. (See \cite{Barvinsky_Vilkovinsky_1990}
for more details.) The series (\ref{HK_4}) if expanded with respect
to the variable $s\Delta$ gives, subject to the remark just made,
the coefficients of $\int\sqrt{g}R_{\mu\nu}\Delta^{n}R^{\mu\nu}$
and $\int\sqrt{g}R\Delta^{n}R$ contributing to the coefficients $B_{2n}$.

We are interested in the $d=2$ case where $R_{\mu\nu}=\frac{1}{2}g_{\mu\nu}R$
and so equation (\ref{HK_4}) becomes: \begin{equation}
\textrm{Tr}\, K(s)=\frac{1}{4\pi s}\int d^{2}x\sqrt{g}\left[1+s\frac{R}{6}+s^{2}R\, f_{R}(s\Delta)R+O(R^{3})\right]\,,\label{HK_6}\end{equation}
where\begin{equation}
f_{R}(x)=\frac{1}{32}f(x)+\frac{1}{8x}f(x)-\frac{1}{16x}+\frac{3}{8x^{2}}f(x)-\frac{3}{8x^{2}}\,.\label{HK_7}\end{equation}
As we mentioned before, the first coefficient in a series expansion
of $f_{R}(x)=\frac{1}{60}+...$ does not match the coefficient $B_{4}$
evaluated in $d=2$ where we find the coefficient $\frac{1}{72}$.
To find agreement between the coefficients, terms of third order in
the curvatures have to be considered in the non-local expansion.

The last technical tool we need in order to evaluate functional traces
are the integrals\begin{equation}
Q_{n}[h]=\int_{0}^{\infty}ds\,\tilde{h}(s)\, s^{-n}\label{HK_8}\end{equation}
that occur when we insert in (\ref{HK_1.1}) the heat kernel expansions
(\ref{HK_2}) or (\ref{HK_6}). Standard result about Mellin transforms
\cite{Codello_Percacci_Rahmede_2009} give the formulas\begin{equation}
Q_{n}[h]=\left\{ \begin{array}{ccc}
\frac{1}{\Gamma(n)}\int_{0}^{\infty}dz\, z^{n-1}h(z) &  & n>0\\
(-1)^{n}h^{(n)}(0) &  & n\leq0\end{array}\right..\label{HK_9}\end{equation}
If in addition a factor $e^{-sa}$ is present in the integral (\ref{HK_8})
then the function $h$ in (\ref{HK_9}) has to evaluated at $z+a$.
For more details refer to Appendix A of \cite{Codello_Percacci_Rahmede_2009}.

\end{document}